**Characterization of Surface and Structure of in-situ Doped Sol-Gel-Derived Silicon Carbide **

By *Olivia Kettner, Sanja Šimić, Birgit Kunert, Robert Schennach, Roland Resel, Thomas Grießer* and *Bettina Friedel\**

[*]    Dr. B. Friedel
illwerke vkw Endowed Professorship for Energy Efficiency, Energy Research Center, Vorarlberg University of Applied Sciences
6850 Dornbirn, Austria
E-mail: bettina.friedel@fhv.at
        Dr. B Friedel, O. Kettner, B. Kunert, Prof. Dr. R. Schennach, Prof. Dr. R. Resel
Institute of Solid State Physics, Graz University of Technology
8010 Graz, Austria
        S. Šimić
Graz Centre for Electron Microscopy (ZFE)
8010 Graz, Austria
        Prof. Dr. T. Grießer
Institute of Chemistry of Polymeric Materials, Montanuniversität Leoben
8700 Leoben, Austria

[**]   O.K. and B.F. are grateful to the Austrian Science Fund (FWF) for financial support under Project No. P 26968.
(Supporting Information is available online from Wiley InterScience or from the author).

***Abstract:*** *Silicon carbide (SiC), is an artificial semiconductor used for high-power transistors and blue LEDs, for its extraordinary properties. SiC would be attractive for more applications, but large-scale or large-surface area fabrication, with control over defects and surface is challenging. Sol-gel based techniques are an affordable alternative towards such requirements. This report describes two types of microcrystalline SiC derived after carbothermal reduction from sol-gel-based precursors, one with nitrogen added, the other aluminum. Characterization of their bulk, structure and surface shows that incorporation of dopants affects the formation of polytypes and surface chemistry. Nitrogen leads exclusively to cubic SiC, exhibiting a native oxide surface. Presence of aluminum instead promotes growth of hexagonal polytypes and induces self-passivation of the crystallites' surface during growth. This is established by* *hydrogenation of silicon bonds and formation of a protecting aluminum carbonate species. XPS provides support for the suggested mechanism. This passivation is achieved in only one step, solely by aluminium in the precursor. Hence, it is shown that growth, doping and passivation of SiC can be performed as "one-pot synthesis". Material without insulating oxide and a limited number of defects is highly valuable for applications involving surface-sensitive charge-transfer reactions, therefore the potential of this method is significant.*

## 1. Introduction

Silicon carbide is a high-temperature stable, non-oxidic ceramic and first been synthesized by A. G. Acheson in 1890,[1] and due to its high Mohs hardness of 9.3, was widely produced and used as an abrasive.[2] Still today, SiC is an important material, used to improve mechanical and thermal properties of machine parts,[3] and preventing from wear in abrasive or corrosive environments.[4-5] But SiC is also a wide band gap semiconductor, highly attractive for high power transistors, light emitting diodes and solar cells, due to its extraordinary electrical properties, often referred to as the third generation of semiconductor materials.[6-9] Silicon carbide appears in more than 200 polytypes, which differ by the stacking sequence of the Si-C tetrahedrals.[10] However, obtaining semiconductor grade material is comparatively more challenging: Single crystals from hexagonal polytypes can be grown via sublimation, where mostly nitrogen-doped or nominally undoped wafers are commercially available.[11] Initially p-doped material is rare because the deeper valence band of wide bandgap materials makes it hard to reach sufficient densities,[12] but can be established post-growth via ion-implantation or CVD epitaxial growth. Cubic polytype 3C-SiC, can merely be grown by heteroepitaxial growth e.g. on silicon,[13] also nitrogen-doped or nominally undoped. However, lattice mismatch and different thermal expansion coefficients lead to mechanical tension in the grown material, with tendency to cracking, therefore 3C-SiC wafers are enormously expensive.[14]

SiC preparation via sol-gel processing is a low-cost, solution-processable and environmentally-friendly addition to those traditional methods,[15] with the potential to expand into further fields of applications particular by its ability to facilitate the production of large-area and high-surface area materials. It is suitable for fabrication of nano- to microscale films, of nano- to microparticles and bulk materials alike.[16] As has been shown, its high degree of freedom regarding precursor composition even enables in-situ doping, such as highly doped n-type SiC:N obtained by the introduction of nitric acid or sodium nitride, and p-type SiC:Al with reasonable charge carrier densities by introduction of aluminum chloride or metallic aluminum powder during wet-chemical precursor preparation.[17-18] Thereby evidence supporting the dopants presence and position in the SiC material was found by electron paramagnetic resonance spectroscopy and associated modelling of the related defect signals for N- and Al-doping, respectively.[16,19] This motivates a broadening of SiC's application range towards new fields, such as the electronic acceptor in organic/inorganic hybrid photovoltaics, as robust photoelectrode in catalysis and for nanostructured blue light emitters, where nanoscale or large surface area is required.[20-21] In particular from hybrid solar cells, it is a well-known fact that the surface properties of the inorganic component substantially determine the efficiency of the device, e.g. via surface trap states fostering trap-mediated recombination or inter-particle charge transport limitations, caused by their stabilizing ligands.[22-24] In the case of silicon carbide nanoparticles within a conjugated polymer matrix, Kettner et *al.* have observed extended emission lifetimes, assigned to the formation of longer lived polarons upon photoinduced electron transfer to the inorganic acceptor or surface trap states,[21] which have been suggested in either case to be beneficial for their function in hybrid photovoltaics.[25] However, also here the surface is the most crucial point, when incorporating SiC nanoparticles into the organic semiconductor matrix of a hybrid solar cell.

Investigations on the surface of traditionally derived SiC have been done in the past, e.g. by Mac Millan et *al.*[26], Afanesev et *al.*[27] and Kaplan.[28] Thereby they mainly focused on the formation of native oxide (SiO$_2$) on hexagonal SiC and the interface formation between SiC and SiO$_2$, also depending on which site was exposed, Si or C. Formation of native oxides on SiC is a strong disadvantage for most (opto)electronic applications, as the insulating SiO$_2$ layer hinders charge transfer. Further, it has been reported that SiC can show considerable densities of surface defects under certain circumstances, for example after etching processes. In one important example, namely blue emitting porous SiC, those defects can detrimentally diminish the original emission properties, and subsequent surface passivation is vital.[29]

In this paper, our experimental observations regarding the dopant-related surface termination of sol-gel derived 3C-SiC crystalline particles are reported. The work aims to give new insights in the special characteristics of surface formation in doped sol-gel derived cubic SiC and, therefore, is a further step to exploit the full potential of this material. The work is based on thorough surface characterization utilizing X-ray photoelectron spectroscopy (XPS) depending on the introduced dopant. This is accompanied by results on their crystallite size, unit cell and polytype distribution, as characterized by scanning electron microscopy (SEM) and X-ray diffraction (XRD) with Rietveld refinement.

## 2. Materials and Methods

### 2.1. Preparation of Silicon Carbide Powders

Silicon carbide sub-micrometer powders were fabricated via sol-gel processed precursors and their carbothermal reduction, which is described in detail elsewhere.[17-18] All chemicals used for the fabrication of the precursor material were purchased from Sigma-Aldrich and used as received without any further purification. In brief, sucrose (> 98.5%) as carbon source and tetraethoxysilane (TEOS) (≥ 99%) as silicon source were dissolved in deionized water and ethanol (≥ 98%), respectively, to get a carbon-rich silica-sol in ethanolic solution as precursor material. The silicon-to-carbon ratio was adjusted to 1:4 and hydrochloric acid (37%) was added as the catalyst. The molar ratio between metal-organic precursor, water and catalyst

TEOS:H$_2$O:HCl is at 1:8.2:27. For p- and n-doping, the precursors were modified by adding nominally 5 at% relative to silicon of aluminum (metallic aluminum powder, 5 µm, 99.5%) or nitrogen (sodium nitrate, ≥ 99%), respectively. The silica-sol was allowed to gel at 60°C in a sealed container for about 12 hours, leading to formation of an amber lyogel, and by drying in an open container at 150°C for 48 hours to get a black xerogel. Next, the precursor was annealed at 1000°C for 3.5 h under argon gas flow, during which left organic degradation products of the former sucrose get carbonized and vitrification of the silicate gel is completed. To convert that gained carbon-rich silicate glass into silicon carbide, the precursor was sintered at 1800°C in argon atmosphere for 15 minutes in an induction furnace. Some nitrogen-doped samples were subsequently annealed in oxygen atmosphere at 1000°C for 10 minutes to demonstrate oxidation effects.

## 2.2. Characterization

Imaging on the SiC crystals was performed via scanning electron microscopy (SEM) (Zeiss DSM 982 Gemini). The microscope was equipped with a thermal field emission gun as source and an Everhart Thornley Detector. Secondary electron images were recorded with an acceleration voltage of 10 keV. The structure was investigated by X-ray diffraction (XRD) using a Siemens D 501 diffractometer in Bragg-Brentano geometry operated at 40 kV and 30 mA using Cu $K_\alpha$ radiation and a graphite monochromator at the secondary side. The XRD data were refined by Rietveld analysis for phase fractions and lattice parameters (Bruker AXS, Topas Version 3.0). Qualitative phase analysis was performed on basis of PDF-2. XPS was performed with a monochromatic Thermo Fisher K-Alpha spectrometer equipped with an Al X-ray source (1486.6 eV) operating at a base pressure in the range of $10^{-8}$ to $10^{-10}$ mbar. High resolution scans were acquired at a pass energy of 50 eV and a step size (resolution) of 0.1 eV. Survey scans were acquired with a pass energy of 200 eV and a step size of 1.0 eV. The instrument work function was calibrated to give a binding energy (BE) of 83.96 eV for the Au 4f$_{7/2}$ line for metallic gold. All measurements were performed at room temperature. The peaks were fitted utilizing Gaussian/Lorentzian mixed functions employing a linear background correction (program XPSPEAK41). Absorption infrared spectroscopy (RAIRS) was used to obtain electronic and chemical material information. A detailed description of the set-up and measurement procedure used for performing reflection measurements can be found elsewhere.[30] In short, the spectra were recorded utilizing a reflection unit with variable angle and a motorized polarizer (Bruker Optics). The measurements were performed with an incidence angle of 74° under vacuum with a base pressure of about 4 mbar. A correction for the substrate absorption (indium foil) was done for the recorded spectra.

## 3. Results

### 3.1. Structural Analysis

Appearance and morphology of the nitrogen- (SiC:N) and aluminum-doped SiC (SiC:Al) powder samples were investigated by SEM. The easiest obtainable form of sol-gel derived SiC (beside fibers, thin films and porous structures) are microcrystals. **Figure 1** shows two examples of microcrystalline samples of SiC:N (a) and SiC:Al (b). Optical microscopy images of the samples (not shown) show SiC:N microcrystals in their typical green colour, while SiC:Al microcrystals appear dark blue. One of the most obvious features of the crystals of SiC:N are the triangular shaped facets, originating from the truncated tetrahedron, typical and characteristic for the zinc blende structure of the cubic SiC polytype 3C-SiC. In comparison, the SiC:Al sample is dominated by hexagonal facets of rather flat crystals, which cannot be clearly assigned to a certain polytype merely from the image. They might still be truncated tetrahedra of different tracht and habit, therefore showing full hexagonal instead of triangular faces and flattened shape. It is another possibility that these are hexagonal flat crystals, which are typical for the so-called Lely platelets of the hexagonal polytype 6H-SiC.

A clear identification and distinction of polytypes in the two doped microcrystalline samples can be obtained by XRD powder diffraction. The XRD patterns of equally derived SiC:N and SiC:Al microcrystalline powders are shown in **Figure 2** (a) and (b), respectively. In the displayed angular range both, the SiC:N (a) and SiC:Al (b) patterns show common most prominent peaks at 35.7°, 41.5°, 60.1°, 71.8° and 75.6°. The according lattice constants were calculated to $d_{111}$ = 2.51 Å, $d_{200}$ = 2.18 Å, $d_{220}$ = 1.54 Å, $d_{311}$ = 1.31 Å and $d_{222}$ = 1.26 Å, respectively. These are in good agreement with literature values for 3C-SiC and can be assigned to its (111), (200), (220), (311) and (222) crystal planes.[31] Further, the 3C-SiC (111) diffraction peak shows two satellites at 34.2° and 38.2° for both samples. With their calculated lattice constants of $d_{101}$ = 2.62 Å and $d_{103}$ = 2.36 Å, they have been identified as the (101) and (103) crystal planes of the 6H-SiC polytype. Thereby the intensity of the satellites in comparison to the 3C-SiC (111) is less intense for SiC:N than for SiC:Al, which could indicate a larger contribution of the 6H-SiC polytype in the SiC:Al sample. The XRD pattern of SiC:N shows further minor peaks, of which all but one could be assigned to 6H-SiC. This confirms that the cubic polytype 3C-SiC and a minor portion of the hexagonal 6H-SiC polytype, dominate the phase composition of the nitrogen-doped SiC microcrystals (Figure 2a). No contributions of $SiO_2$ or foreign species such as silicon nitride were found. In comparison, the pattern for the aluminum-doped SiC microcrystals (Figure 2b) is slightly more complex. Here, the 3C-SiC phase is still the most prominent one, according to their peak intensity, but overlaid with a large number of further diffraction peaks. Part of them are the aforementioned 6H-SiC satellites at 34.2° and 38.2°, along with the additional minor signals of 6H-SiC, which in general show higher intensity than for the ones in the SiC:N powder. Two small shoulders to the 6H-SiC peaks appearing in both sample patterns at 37.7° and 64.7°C are caused by negligible contributions of the polytype 15R-SiC. The additional diffraction peaks, which are not present for SiC:N, have been assigned to the hexagonal polytype 4H-SiC. It should be noted that the peak at 26.6° is a $SiO_2$ contribution (quartz (011)), does not originate from of the actual sample (as confirmed by FTIR and XPS) but is an external contamination with agate. No foreign impurities such as aluminum oxide have been found. These structural results indicate that the SiC:Al samples exhibit a different polytype composition than SiC:N despite identical preparation conditions besides the dopant.

For a distinct quantitative analysis of the phase-composition of the samples, Rietveld refinement on the respective XRD patterns has been performed to obtain the dopant-dependent unit cell dimensions and the crystallite sizes of certain polytype phases.[32-33] The obtained values are summarized in **Table 1** and **Table 2** for SiC:N and SiC:Al, respectively. The detailed comparison of the observed pattern with the data gained from Rietveld analysis and their differential-plots can be found in the supplementary information. According to Rietveld refinement, the SiC:N powder consists of 93% of the cubic 3C-SiC polytype and 7% of the hexagonal 6H-SiC. The peaks belonging to 3C-SiC are sharp and can be fitted to crystal sizes larger than 200 nm. The peaks assigned to the 6H-SiC are broadened, leading to calculated crystal sizes of around 30 nm. There is no indication of other polytypes in the SiC:N sample according to Rietveld refinement. In comparison, the evaluation of data collected from the SiC:Al microcrystalline powder revealed a phase-composition of about 53% 3C-SiC, 37% 6H-SiC and 10% 4H-SiC polytype. In this case, 3C-SiC and 6H-SiC both, gave particle sizes larger than 100 nm, while the line broadening of peaks related to 4H-SiC indicate smaller particle sizes of around 30 nm. This confirms first indications from SEM imaging that SiC:N indeed is dominated by the cubic 3C-SiC polytype of relatively large sizes, while the smaller particles can be assigned to the hexagonal 6H-SiC minority. The SiC:Al on the other hand has large contributions of both, the cubic 3C-SiC and the hexagonal 6H-SiC polytype with intermediate crystal sizes, while here a minority of small particles of another hexagonal polytype 4H-SiC have formed. A reason for the difference in polytype formation despite identical synthesis parameters must be related to the presence of the dopant. Thereby the effect of the dopant on polytype formation can have a kinetic as well as a chemical origin, as it was shown by Jepps et *al.* that nitrogen as impurity promotes the growth of 3C-SiC.[10] Aluminum, in contrast, tends to stabilize the hexagonal modifications.[34]

The unit cell data (axial length *a* for cubic, basal axial length *a* and height *c* for hexagonal unit cells) of each polytype fraction of the SiC:N and SiC:Al samples were derived by Rietveld refinement and are also presented in Tables 1 and 2, respectively. The value for the lattice constant of nominally undoped 3C-SiC at room temperature is specified with $a_{3C\text{-}SiC}$=4.3596 Å,[35] which is in good agreement with the obtained unit cell data for the cubic component of SiC:N with $a_{3C\text{-}SiC:N}$ = (4.359±0.005) Å and SiC:Al with $a_{3C\text{-}SiC:Al}$ = (4.36±0.01) Å. Regarding hexagonal contributions the polytypes 6H-SiC and 4H-SiC are significant for Rietveld refinement of the samples SiC:N (6H) and SiC:Al (6H, 4H). Thereby literature refers to lattice parameters of basal axis of $a_{6H\text{-}SiC}$ = 3.0806 Å and height of $c_{6H\text{-}SiC}$ = 15.1173 Å for nominally undoped 6H-SiC.[35] The calculated values for SiC:N are $a_{6H\text{-}SiC:N}$ = (3.081±0.005) Å and $c_{6H\text{-}SiC:N}$ = (15.118±0.005) Å, and for SiC:Al $a_{6H\text{-}SiC:Al}$ = (3.09±0.01) Å and height of $c_{6H\text{-}SiC:Al}$ = (15.13±0.01) Å, respectively. All parameters lie entirely within this range, merely the height of the 6H-SiC:Al unit cell appears to be slightly elongated, even with the considered deviation of the refinement. Reported unit cell dimensions for 4H-SiC are $a_{4H\text{-}SiC}$ = 3.0730 Å for its axis and $c_{4H\text{-}SiC}$ = 10.0530 Å for its height.[2] Rietveld refinement on the SiC:Al here led to $a_{4H\text{-}SiC:Al}$ = (3.08±0.01) Å and height of $c_{4H\text{-}SiC:Al}$ = (10.10±0.01) Å, again in good agreement for the axis component, but a significant elongation of the height. Comparing these lattice parameters of sol-gel derived nitrogen- and aluminum-doped SiC with literature, it can be observed that the hexagonal unit cells for SiC:Al are slightly larger, while they are unchanged for the cubic component or nitrogen incorporation. The elongation of lattice parameters for hexagonal SiC:Al, which makes $\Delta c_{6H\text{-}SiC:Al}$ = (0.01±0.01) Å and $\Delta c_{4H\text{-}SiC:Al}$ = (0.05±0.01) Å can be explained by the larger covalent radius of the Al atom (1.18 Å), which resides on a silicon-position (Si covalent radius 1.11 Å) in the lattice, which has been proven by Li et *al.* for 4H-SiC.[36] Nitrogen, which resides at carbon sites,[37] has a slightly smaller covalent radius of 0.75

Å, compared to carbon with 0.77 Å, but does not lead to significantly changed unit cell dimensions in this case.[38]

### 3.2. Surface Composition

Surface chemistry of the differently doped sol-gel derived SiC powders was investigated via XPS to identify any connections between the surface termination and the incorporated dopant, in particular the effect on native oxide formation. Therefore the silicon *Si 2p* and carbon *C 1s* core level spectra were utilized to distinguish different silicon and carbon species in the samples and are shown in **Figure 3**. A comparison of the *Si 2p* core level spectra of SiC:N (top) and SiC:Al microcrystalline powder (bottom) is shown in Figure 3a. The *Si 2p* spectra can be deconvoluted into peaks arising from the different oxidation states of silicon $Si^0$, $Si^{1+}$, $Si^{2+}$, $Si^{3+}$ and $Si^{4+}$,[39] which are located at different binding energies, thus allow identification of chemical bonds and make deductions of related compounds. In the present case, the *Si 2p* spectrum of SiC:N shows only two distinct peaks, one at a binding energy of 103.7 eV and another at 100.8 eV, whereas the latter has considerably higher intensity. While the low-energy signal can be assigned to the $Si^+$ state in SiC, [40] the other at higher energy arises from $Si^{4+}$ in $SiO_2$, [41-43] indicating the presence of a significant native surface oxide layer on the SiC:N microcrystals. In comparison, the *Si 2p* core level spectrum of the SiC:Al microcystals shows merely one dominant peak at a binding energy of 99.7 eV with a tail towards the high-energy side. The strong peak can be assigned to $Si^0$, which originates either from Si-Si bonds in bulk silicon, [44-45] or from Si-H bonds, as typically observed for hydrogenated SiC at this energy.[46-48] The tail originates from a quite weak underlying peak centered at 101.3 eV, which can be assigned to a shifted $Si^+$ signal from SiC or a $Si^{2+}$ contribution originating from an imperfect sub-oxide layer ($SiO_x$ with $0<x<2$) or a combination of both.[39,49] There are no signs of a native oxide layer on the surface of the SiC:Al crystallites.

The *C 1s* core level spectra of SiC:N (top) and SiC:Al (bottom) microcystalline powders are shown in Figure 3b. The SiC:N spectrum exhibits an asymmetric line feature built by two underlying peaks, a strong one located at 282.9 eV and a minor one at 284.8 eV. The low-energy peak at 282.9 eV can be clearly identified as originating from the C-Si bond of silicon carbide.[43,50] The less intense peak at 284.8 eV can roughly be assigned to some kind of C-C bond, but the exact origin is unclear.[50] In literature, *C 1s* peaks occurring at 285.0 eV ±0.4 eV have been discussed related to various origins,[51] such as diamond-like $sp^3$ carbon (285.0 eV),[52] graphitic $sp^2$ carbon (284.6 eV), [53] adventitious carbon (285.1 eV) from organic post-fabrication deposites,[54] or mixed $sp^2/sp^3$ phases.[55] Estrade-Szwarckopf investigated asymmetric *C 1s* features, where a broad peak occurred, shifted against the graphitic $sp^2$ peak with its intensity varying with the treatment of the sample and referred to this broad 285 eV peak as "defect carbon".[56] Also Emtsev et *al.* and Rani et *al.* found such a broad *C 1s* peak close to 285 eV when investigating graphene grown epitaxially onto SiC or Gold, respectively, and also ascribed it to $sp^2$ defects at the graphene-substrate interface.[57-58] Finally, Iwanowski et *al.* suggested carbonaceous surface exclusions from a carbon-saturated cubic silicon carbide ($Si_{1-x}C_x$ with x>0.54) crystal lattice, being responsible for the observed 285 eV peak. [59] In the present *C 1s* spectra for SiC:N, the small signal intensity of that peak makes surface $sp^2$, $sp^3$ or mixed carbons rather unlikely as origin, and also adventitious carbons can be excluded in absence of exposure to organics between high-temperature synthesis and characterization. Therefore, here it is suggested that the observed 284.8 eV feature is related to buried defect C-C bonds at the interface between the SiC and the native silicon oxide layer.[60]

The *C 1s* spectrum of the SiC:Al microcrystalline powder indicates at first glance a completely different surface composition in comparison to SiC:N. The spectrum can be deconvoluted into two almost equally strong peaks at 281.9 eV and 284.5 eV and a very weak one at 289.1 eV. The most intense feature, arising at 281.9 eV, was attributed to C-Si bonding in silicon carbide.[44] Its significant shift in binding energy compared to SiC:N can be ascribed to the notably different sample composition, because binding energies are known to shift with the polytype.[59] The slightly less intense peak with its maximum at 284.5 eV was clearly identified as C-C bonding of graphitic surface carbon.[61] The weak signal at 289.1 eV was ascribed to O-C=O bonding, suggesting the formation of minor carbonates on the surface.[62] These results indicate that the surface structure of the SiC:Al crystallites is composed of a hydrogenated silicon interface buried under a graphene surface layer, similar to what has been reported by Pallechi et al..[63] Thereby the thickness of the graphene layer can be estimated from the intensity ratio between the carbide and graphitic peak to be nominally between one and two monolayers.[64] The very small carbonate signal suggests that the $Al_2(CO_3)_3$ is merely present as a small density of distributed "defect-like" sites below or above the carbon surface.

### 3.3. Bulk Properties

The RAIRS spectra of SiC:N and SiC:Al microcrystals are illustrated in **Figure 4**. Both samples clearly show a strong peak between 750 and 1000 cm$^{-1}$, which is referred to as the "*reststrahlen band*", a typical feature arising from the phonon-polariton resonance of bulk SiC. This peak generally dominates spectra recorded from SiC and is caused by vibrations of the Si- and C-sublattices against each other.[65] The tail of the phonon band towards higher wavenumbers for SiC:N is an effect that originates from high dopant concentration.[66] The unusual sharpness of the peak compared to the rather broad reststrahlen band usually seen for SiC single crystal wafers is a result of the present microcrystalline morphology of the material. Narrowing of the reststrahlen band has been reported for porous structures and powders and assigned to an increase in surface area.[26] On the high wavenumber side, where SiC:Al shows no further features, SiC:N exhibits an additional absorption band at about 1110 cm$^{-1}$. This peak can be assigned to Si-O-Si vibrations of $SiO_2$.[67] This peak is superposed with a very broad absorption in the region between 2350 and 970 cm$^{-1}$. The latter is a result of plasmon-phonon interactions, which occur when plasmon and phonon modes appear in a similar frequency range.[65] The plasma frequency of undoped SiC lies typically in the microwave range, but strongly depends on charge carrier concentration, effective electron mass and the electron mobility. Here, the shift to higher energies emerges from large charge carrier concentration, which was found to be $N_D-N_A = 2.1 \cdot 10^{19} cm^{-3}$ from Hall-measurements on this synthesized material.[17] The small absorption peak at about 640 cm$^{-1}$ for SiC:N can be attributed to the coupling of the longitudinal optical phonon to the plasmon.[65] The absence of both these features in SiC:Al is caused by its considerably lower charge carrier concentration from p-doping, which was found to be merely $N_D-N_A = -6.8 \cdot 10^{11} cm^{-3}$.[68] The absence of the 1110 cm$^{-1}$ oxide peak in the spectrum of the SiC:Al is in good agreement with the XPS results, which is showing no silicon oxide present on the surface. The absence of a carbonate contribution in spectrum is not surprising because it appeared only in trace amounts on the surface, and undetectable for rather bulk sensitive IR measurements.

## 4. Discussion

While the incorporation of a nitrogen dopant with this SiC synthesis method results in SiC:N microcrystals exhibiting typical common surface composition of SiC with a native oxide (SiO$_2$) layer and some imperfect graphite species, the equivalent incorporation of aluminum in the process leads to a very different surface composition. The SiC:Al exhibits apparently a quite stable hydrogen-terminated surface (Si-H), most likely buried under a graphene mono- or bilayer surface, while oxygen is merely found bound in a very small density of aluminum carbonate sites near the surface. These seem to cause a self-passivation effect on the SiC:Al crystallites and hinder formation of a native oxide layer. Hydrogen is known to saturate dangling bonds in semiconductors, especially in silicon, but also in SiC and therewith deactivating potential luminescence quenching or charge trapping sites.[29] The presence of a high density of Si-H bonds and absence of a Si-dangling bond signal (BE < 99eV) for SiC:Al here, indicates successful deactivation of these defects. Further, the carbonate layer, despite having less than nominally 1nm thickness judged by the intensity of the signal, seems to have passivated the SiC surface against the typical formation of silicon oxide at exposure to air. Potential reasons for this different surface composition of the present SiC:Al lay in the fact that during this "one-pot synthesis" and in particular during the final annealing step, all chemical components are enclosed in the reaction chamber volume. The carbothermal reduction of silicon species in the precursor and formation of SiC occurs merely at temperatures > 1700°C, following the reaction $SiO_2$ (s) + 3C (s) → SiC (s) + 2CO (g). This happens in two ways: (1) directly reacting with solid carbon via $SiO_2$ (s) + C (s) → SiO (g) + CO (g) and SiO (g) + 2C (S) → SiC (s) + CO (g), and indirectly by reaction with CO via $SiO_2$ (s) + CO (g) → SiO (g) + $CO_2$ (g) and SiO (g) + 3CO (g) → SiC (s) + $2CO_2$ (g). As also the aluminum compounds go into the gas phase during this process, where some aluminum is built into the SiC lattice as dopant, also secondary chemical reactions between Al and $CO/CO_2$ are likely. It is suggested that here aluminum carbonate is formed at the surface of the SiC crystals during the cooling below 1700°C, when silicon species are no longer in a stable gas phase nor reacting with C or CO. Three scenarios are imaginable for their formation: (1) Few separated near-surface aluminum sites in the SiC lattice react with the $CO/CO_2$ gas; (2) remaining gaseous aluminum species condensate on the SiC surface and subsequently react with surface carbon or the $CO/CO_2$ gas; (3) gaseous aluminum species react in the gas phase with $CO/CO_2$ and subsequently condensate on the SiC surface. However, since aluminum carbonate $Al_2(CO_3)_3$ is known not to be environmentally stable, thus would have been decomposed to $Al(OH)_3$ by the time of measurement. Therefore it is assumed that the carbonate sites are also buried underneath the graphene surface layer, which stabilizes them. No other aluminum species like $Al_2O_3$ or $Al(OH)_3$ were found. This was also confirmed by FTIR, which also proves the absence of $SiO_2$ in the SiC:Al sample and further demonstrates the typical difference in charge carrier concentration between the n- and p-doped SiC by the appearance of the plasmon-phonon interaction. The observed formation of different polytypes, triggered by the incorporated dopant and also the different sizes of the according crystallites, allows certain speculations about the role of the dopant in the crystal growth process. The dopant might be acting "only" as a catalyst or by its displacement of carbon or silicon in the lattice, leading to changes of the unit cell size and thus giving preference to certain polytype formation. The role of the dopant as chemical pathway with strong impact on the final polytype has also been discussed by Ariyawong et *al.* who addressed the link between the crystal chemistry and growth process parameters.[69] Therein SiC is treated as a solid solution and the polytypes evaluated in terms of their respective C and Si activities. They found that 3C-SiC is always obtained with high Si activity in SiC, i.e close to the SiC-Si two-phase boundary. In the present case, the dominant polytype for SiC:N is 3C-SiC and the crystals are oxidized on the surface. With Al incorporation in SiC:Al, the hexagonality increases and the SiC surface is not oxidized. Therefore it could be considered that the reducing conditions on SiC by Al might increase the C activity in the crystals, thus promoting formation of the hexagonal polytypes.

Most important in terms of usability of this material in electrochemical and (opto)electronic applications is clearly the role of aluminum for self-passivation of the as-formed crystals, protecting from oxidation and most likely deactivation of surface defects by saturation with hydrogen. The deactivation of dangling bonds and prevention of insulating oxide layers is vital for efficient charge transfer and transport in this material e.g. in LEDs or catalysis. This sort of surface passivation is usually only achieved by post-treatment such as alumina or titania deposition on freshly HF-etched SiC surfaces.[29] In contrast, our synthesis route via sol-gel precursors and carbothermal reduction allows to obtain SiC growth, doping and surface passivation as a one-step process.

## 5. Conclusions

In summary, nitrogen- and aluminum-doped SiC microcrystalline powders have been synthesized from sol-gel based precursors after carbothermal reduction and have been investigated regarding the effect on surface properties and structural changes triggered by the incorporated dopant. Therefore SEM, XRD, XPS and RAIRS were used to gain knowledge about the polytype composition, crystallite size, unit cell size, bulk chemistry and surface chemistry.

It was revealed that SiC:N consists mostly of the cubic polytype 3C-SiC, while SiC:Al is devided into 50% 3C-SiC and 50% hexagonal polytypes 6H-SiC and 4H-SiC. This might be caused either by catalytic properties of the aluminum, promoting formation of the more energy expensive formation of hexagonal phase, or triggered by the change in unit cell size, which decreases where the smaller nitrogen atom replaces carbon and increases where the larger aluminum atom replaces silicon. Independent from this bulk effect, also the surface composition of the two materials is very different. While the nitrogen dopant leads to formation of SiC:N microcrystals with a typical $SiO_2$ surface layer, the equivalent incorporation of aluminum generates SiC:Al crystallites with a surface of hydrogenated silicon bonds and a small density of aluminum carbonate sites buried under a graphene layer. This leads to two valuable effects for SiC: deactivated defect sites by hydrogen saturation of dangling bonds and passivation of the crystallites surface to prevent formation of a native oxide.

The SiC:N instead exhibits no signs of hydrogenation and no graphitic carbon, but the commonly seen native oxide layer, where dangling bonds are saturated by unfavorable oxygen. This difference between SiC:N and SiC:Al might be caused by the high reactivity of aluminum and the longer stability of gaseous aluminum species after SiC formation at the end of the high-temperature process, leading to the natural formation of a passivation layer on the SiC crystallites directly after growth. The nature of this passivation layer appeared as a carbonate feature in XPS, which for aluminum is known to be unstable in air. Therefore it is suggested that the carbonate is formed at the interface between the SiC and the graphene surface, which stabilizes it. The fact that SiC growth, doping and surface passivation can be achieved in one step with our „one-pot synthesis" is a vital key for the application of such affordably prepared materials in applications involving surface-sensitive charge-transfer reactions. This is promising for its potential use in LEDs, photovoltaics and catalysis.

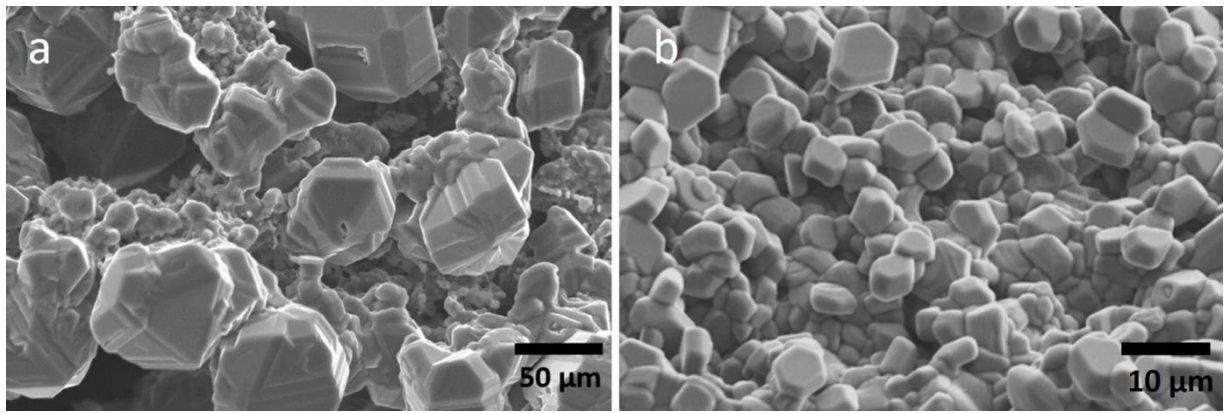

*Fig. 1. SEM images of doped silicon carbide microcrystals as derived from the sol-gel based precursor after carbothermal reduction for a nitrogen- (a) and aluminum-doped SiC sample (b).*

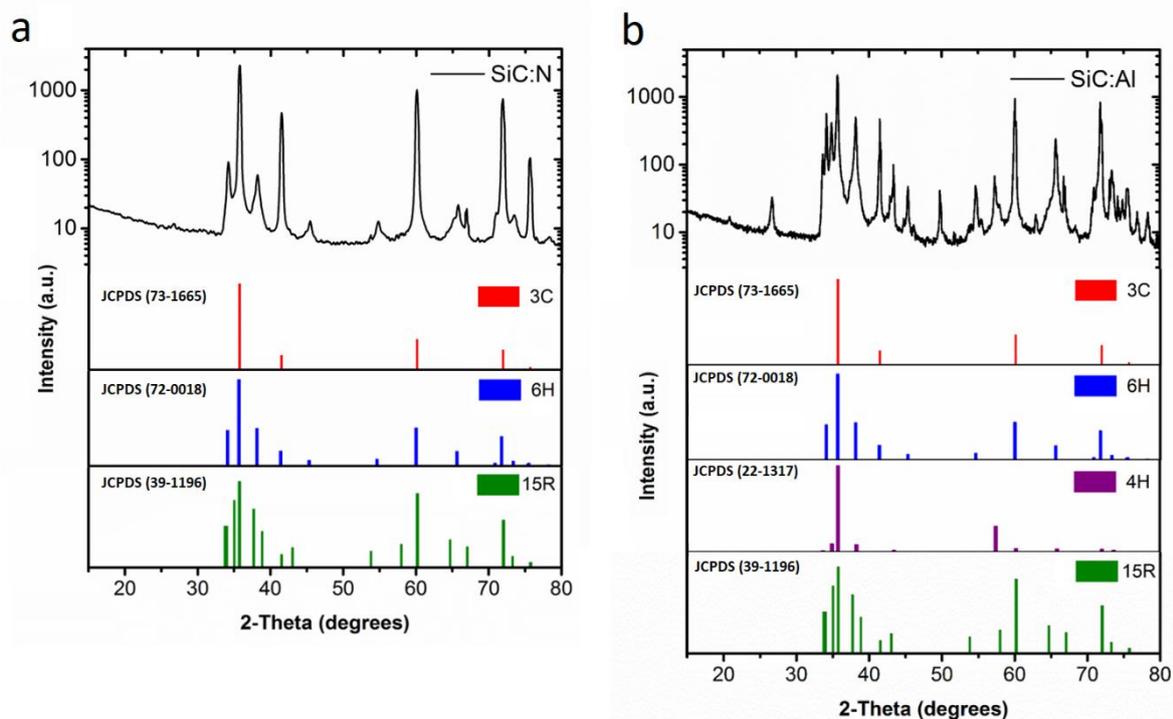

*Fig. 2. X-ray diffraction patterns of pristine nitrogen-doped SiC powder (a) and aluminum-doped SiC powder (b) in comparison with the reference pattern of the different SiC polytypes 3C (red bars, JCPDS 73-1665), 6H (blue bars, JCPDS 72-0018), 4H (purple bars, JCPDS 22-1317) and 15R (green bars, JCPDS 39-1196). (Note: The peak at 66.9° is a machine-related artefact, the peak at 26.6° is a preparation-induced agate contamination).*

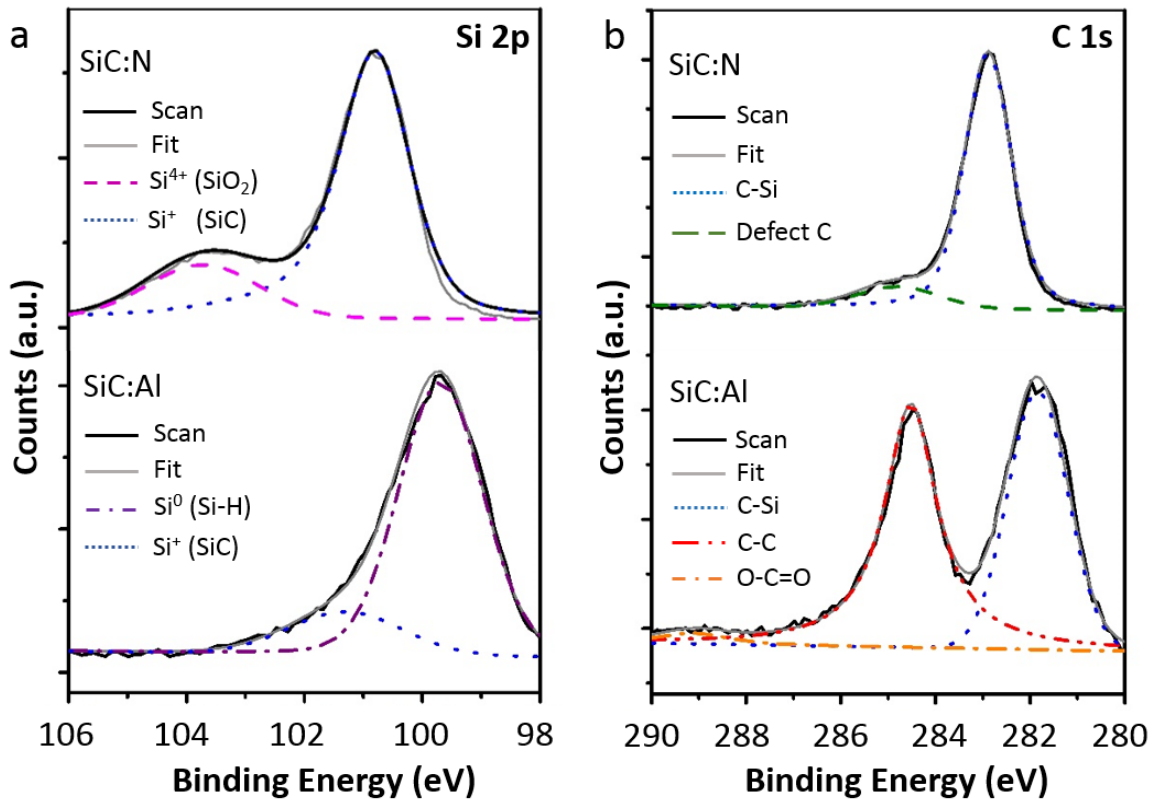

*Fig. 3. XPS spectra (solid line) for Si 2p core level (a) and C 1s core level (b) with simulated peak deconvolution of the components (broken lines) for SiC:N (top) and pristine SiC:Al (bottom) microcrystalline powders. For better visibility, spectra have been normalized.*

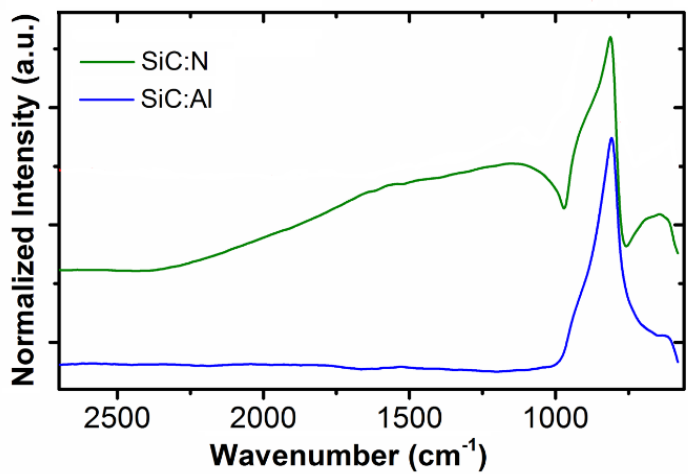

*Fig. 4. Infrared reflection absorption (RAIRS) spectra of SiC:N (top curve) and SiC:Al (bottom curve) microcrystalline powders. For better comparability, spectra have been normalized.*

*Table 1. Phase composition, unit cell dimensions and crystallite sizes determined by Rietveld for a nitrogen-doped SiC powder.*

| SiC Polytype | Composition [%] | a [Å] | c [Å] | Particle size [nm] |
|---|---|---|---|---|
| 3C | 93 | 4.3585 | - | 200 |
| 6H | 7 | 3.0807 | 15.1174 | 30 |

*Table 2. Phase composition, unit cell dimensions and crystallite sizes determined by Rietveld for an aluminum-doped powder.*

| SiC Polytype | Composition [%] | a [Å] | c [Å] | Particle size [nm] |
|---|---|---|---|---|
| 3C | 53 | 4.3603 | - | 100 |
| 6H | 37 | 3.0817 | 15.1275 | 100 |
| 4H | 10 | 3.0800 | 10.0980 | 30 |